\documentstyle[prl,twocolumn,aps]{revtex}
\input{epsf}
\begin{document}
\draft
\twocolumn[\hsize\textwidth\columnwidth\hsize\csname @twocolumnfalse\endcsname
\title{To the problem of Poincar\'e recurrences in generic Hamiltonian systems}

\author{B. V. Chirikov$^{(a)}$  and D. L. Shepelyansky$^{(b)}$}

\address {$^{(a)}$Budker Institute of Nuclear Physics, 
630090 Novosibirsk, Russia \\
$^{(b)}$Laboratoire de Physique Quantique, UMR 5626 du CNRS, 
Universit\'e Paul Sabatier, F-31062 Toulouse Cedex 4, France}

\date{July 30, 2002}

\maketitle

\begin{abstract}
We discuss the problem of Poincar\'e recurrences 
in area-preserving maps and the universality of their decay at long times.
The work is related to to the results presented in Refs. {\protect [1,2]}.
\end{abstract}
\pacs{PACS numbers: 05.45.Mt}
\vskip1pc]

\narrowtext

The authors of Comment  \cite{comment} question the existence
of universal power-law decay of Poincar\'e recurrences 
discussed in our Letter \cite{letter} for generic Hamiltonian systems.
There are two aspects in their quarry:

The first one 
states that the asymptotic decay $P(\tau) \propto1/\tau^3$, which is
expected from the scaling theory of universal phase-space structure
in the vicinity of critical golden curve 
(see Refs. 1-2, 12-18 in \cite{letter}),  is not valid.
Such a conclusion is based on Fig. 1 in \cite{comment} obtained from
numerical  simulations of the standard map with the total number
of map iterations $N_{tot}$ being only by a factor 10 larger than 
for the similar case of Fig. 2 in \cite{letter} where $N_{tot}=10^{12}$. 
New data for $8 < \log \tau \leq 9$ indeed demonstrate 
noticeable deviations from the theoretical estimates for $P(\tau)$
obtained from the exit times $\tau_n$ from golden resonance scales $r_n$
(Fig. 1 in \cite{letter}). These deviations are rather intriguing.
Indeed, the local properties of critical golden curve 
are known to be self-similar and universal (e.g. phase-space structure,
size of resonances and local diffusion rate, see e.g. 
Refs. 2, 15 in \cite{letter}).
Moreover, the exit times $\tau_n$, found in \cite{letter} for a first time, 
give an example of a non-local characteristic
which follows the universal scaling law. The arguments based on this scaling 
lead to the asymptotic decay $P(\tau) \propto 1/\tau^3$ which
however can start after extremely long times $\tau > \tau_{a}$.
In \cite{letter} we have shown that at least 
$\tau_a > \tau_g \approx 2 \times 10^5$
and it is not excluded that this time scale is still much longer
\cite{note}, and is
not yet reached even
in the simulations presented in \cite{comment}.

If to assume this then there is an interesting possibility to see
if $P(\tau)$ would have some universal 
properties on the presently available (intermediate) time scales 
($\tau \leq 10^9$).
This is the second aspect of the quarry on which the authors of the Comment
tend to give a negative answer. To this end we show
in Fig. 1 an example of another map with critical golden curve which
power-law decay $P(\tau) \propto 1/\tau^p$ has the exponent
$p \approx 1.5$. This average value of exponent $p$ 
in the range $10^2 < \tau < 10^8$
agrees, indeed, with that of the 
decay in the standard map in the range $10^5 < \tau < 10^9$
shown in  \cite{comment}. 
There is also a clear similarity for the decay of $P(\tau)$
in two maps (see Fig. 1).
For other values of parameter $\lambda$ the decay $P(\tau )$
in the separatrix map is no longer a simple power law but rather some
irregular oscillations around that which represent a peculiar phenomenon of
the so-called renormalization chaos 
(Ref. 12 in \cite{letter}).
However, averaging over several $\lambda$ values smoothes away the signs of 
renormalization chaos and reveals the underlying picture of the power-law
distribution with the same exponent $p \approx 1.5$ 
(see Ref. 9 in \cite{letter} and \cite{com1}).
Certainly, further studies are required for the understanding of the origin 
of this intermediate
asymptotics and for the estimates of the time scale $\tau_a$ after which
a transition to $p=3$ is expected.

\begin{figure}
\epsfxsize=3.20in
\epsfysize=2.86in
\epsffile{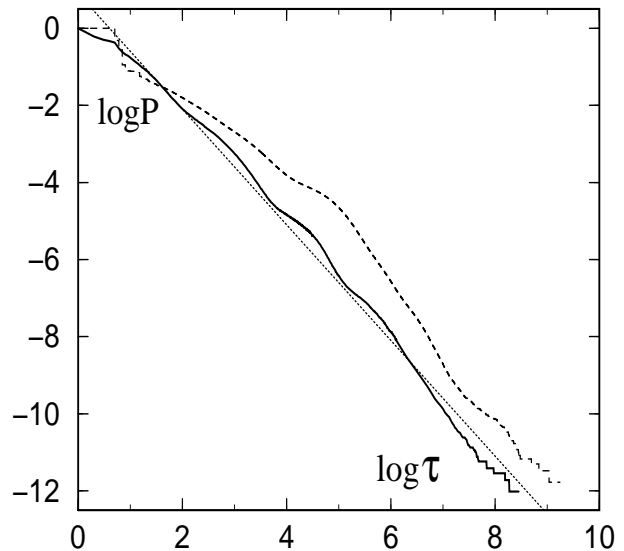}
\vglue 0.2cm
\caption{Poincar\'e recurrences $P(\tau)$ in the separatrix map
(${\bar y} =y +\sin x, {\bar x} = x - \lambda \ln |{\bar y}|$)
with critical golden boundary curve
at $\lambda = 3.1819316$  (return line $y=0$, average return time 
$<\tau> \approx 10$). 
Data are obtained from 10 orbits computed for $10^{12}$ map iterations
(solid curve). 
Dashed curve shows $P(\tau)$ for the standard map at $K=K_g$ 
(data of the lower curve of Fig. 1 in [1]).
The dotted straight line
shows the power-law decay $P(\tau) \propto 1/\tau^p$
with $p=1.5$; logarithms are decimal.}
\label{fig1}
\end{figure}


\begin{thebibliography}{99}
\bibitem{note} The measure of a sticking region is of the order
        of  $\mu(\tau) \sim \tau P(\tau)/<\tau>  \sim 1/q^2$
        and therefore even for $\tau \sim 10^9$ (Fig. 1 in \cite{comment})
        only resonances with $q \sim  100$ are reachable for
        an orbit.
\end{thebibliography}
\end{document}